\documentclass[aps,prb,showpacs,preprint,tightenlines]{revtex4}     
\usepackage{bm}
\usepackage{graphicx}
\begin{document}

\hfill Preprint, in press APL

\title{Theoretical analysis of spectral gain in a THz quantum cascade laser:
prospects for gain at 1 THz}
\author{S.-C.~Lee}
\author{A.~Wacker}
\affiliation{Institut f\"{u}r Theoretische Physik, 
Technische Universit\"{a}t Berlin, Hardenbergstra{\ss}e 36, 10623 Berlin,
Germany}

\received{\ \ \ \ \ \ \ \ \ \ }

\begin{abstract}

In a recent Letter [Appl. Phys. Lett. {\bf 82}, 1015 (2003)], 
Williams {\it et al.} reported the development of a terahertz 
quantum cascade laser operating at 3.4 THz or 14.2 meV. 
We have calculated and analyzed the gain spectra of the 
quantum cascade structure described in their work, and in addition to gain 
at the reported lasing energy of $\simeq$ 14 meV, we have 
discovered substantial gain at a much lower energy of around
5 meV or just over 1 THz. This suggests an avenue for the development
of a terahertz laser at this lower energy, or of a two-color terahertz laser.
\end{abstract}
\pacs{78.67.-n, 42.55.Px, 05.60.Gg}

\maketitle

Following on the initial successful development of quantum
cascade  lasers in the mid-infrared,\cite{FaiGma}
the field has continued
its advance with the recent appearance of terahertz (far-infrared) quantum
cascade lasers. The initial 
examples\cite{KohRoc} of these terahertz lasers, operating
at $\simeq 4.5$ THz, were based on chirped superlattice structures.
An alternative design, operating at $3.4$ THz, and
 based on a simple four well (per period)
structure was recently reported\cite{Wil03} by Williams {\it et al.}
The simplicity of this structure provides a testbed
for a nonequilibrium Green's function 
theory we have recently developed\cite{Lee02a,Wac02b}
to determine the nonequilibrium stationary state
of quantum cascade laser structures operating under an applied
voltage. This theory enables us to analyze transport and gain
properties of these structures, giving us a tool to
evaluate current-voltage (I-V) characteristics and nonequilibrium
distribution functions, 
and to estimate level populations and lifetimes. In addition, we can
use these parameters as a basis for evaluating and analyzing
gain and absorption spectra in these structures. We have applied
this theory to the structure reported\cite{Wil03} by Williams
{\it et al.}, and we report here the results of our theoretical
analysis. We highlight especially the new finding of substantial
gain at the low energy of  $\simeq 5$ meV, well below the spectral
region considered in Ref. \onlinecite{Wil03}.

\begin{figure}
\includegraphics[height=5.5cm,keepaspectratio]{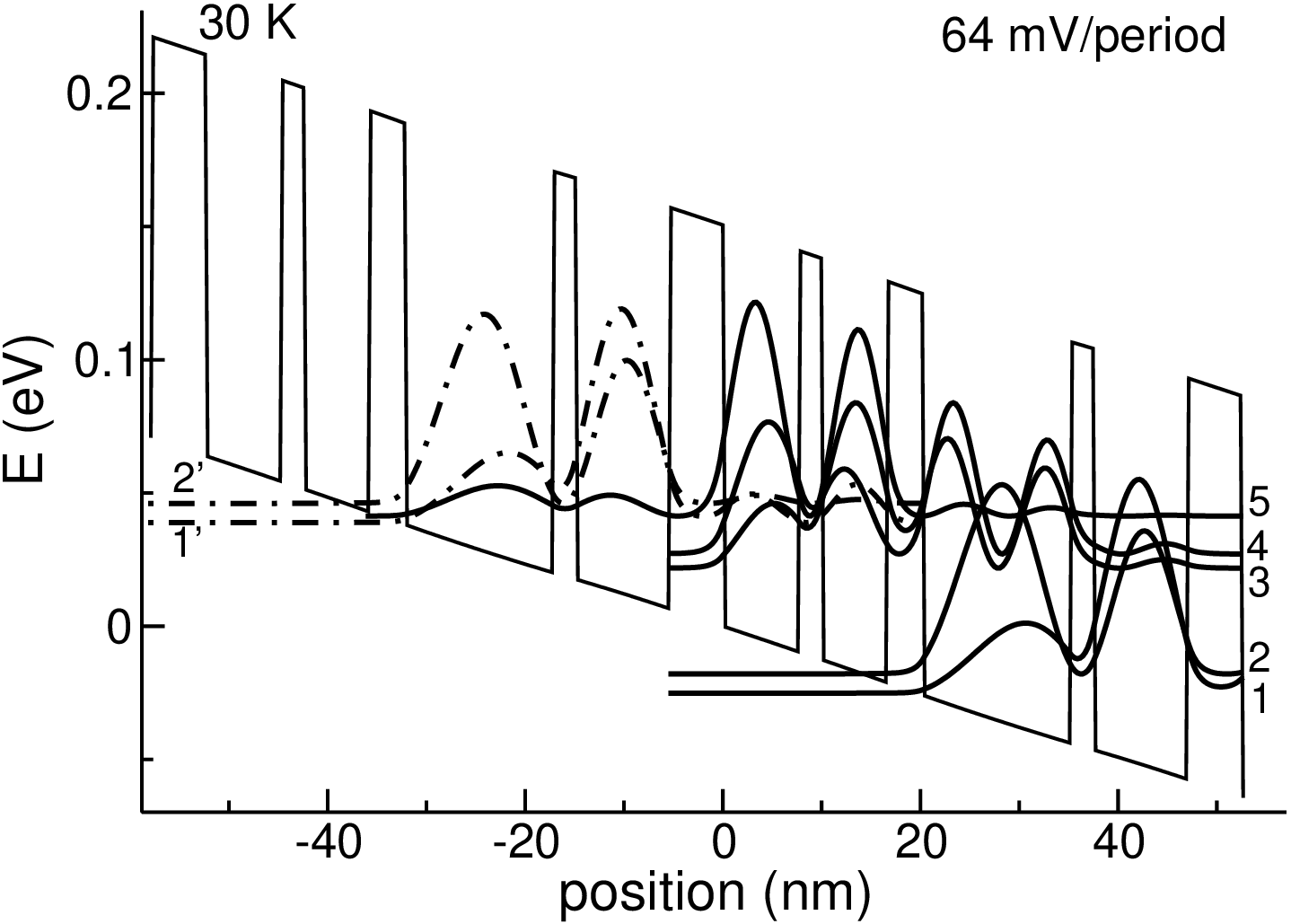}
\caption{Conduction band profile of structure reported in
Ref.~\onlinecite{Wil03} with an applied voltage of 12.2 kV/cm,
and Wannier-Stark levels. Wavefunctions 
(modulus-squared) are also shown.}

\label{fig.potapl}
\end{figure}

Figure \ref{fig.potapl} shows the conduction band profile of two periods
of the structure 
reported in Ref. \onlinecite{Wil03} with an applied bias of 64 mV/period or 
12.2 kV/cm. The Wannier-Stark wavefunctions (modulus-squared) are also shown.
The energetic positions of the wavefunctions shown in the
figure are the Wannier-Stark 
energies renormalized by the mean-field due to electron-electron
scattering.\cite{Lee02a}  Following Ref. \onlinecite{Wil03}, 
we designate levels 4 and 5 as the lower and upper laser levels respectively. 
In addition, we label level 1 (and 1' for the neighboring period) 
the lower collector level, and level 2 (and 2') the upper collector level.

We calculated\cite{Lee02a} the gain spectra by considering a simple two-level
model interacting with a classical time-dependent electric field
which gives the gain coefficient\cite{Hau93}
\begin{equation}
g(\omega) \; = \; -\frac{\omega\pi}{cn_B\epsilon_0 L_p}
\sum_{{ij \atop (E_j > E_i)}}|d_{ij}|^2 (n_i - n_j)\, \mathcal{L}_{ij}(\omega),
\label{eq.chi2lev}
\end{equation}
where we sum over contributions from each pair of levels $i$ and $j$.
$L_p = 52.4$~nm is the length of one period of the structure, 
and $n_B = \sqrt{13}$ is 
the background refractive index.
The Lorentzian $\mathcal{L}_{ij}(\omega)  = 
(\Gamma_{ij}/2\pi)/[(\hbar\omega - \Delta E_{ji})^2
+ (\Gamma_{ij}/2)^2]$, with $ \Gamma_{ij} = \Gamma_i + \Gamma_j$, and
$\Delta E_{ji} = E_j - E_i$. The populations, e.g.,  $n_i$, 
 and the broadening parameters, e.g., $\Gamma_i$, for
level $i$ are extracted
from the nonequilibrium Green's functions as described in 
Ref.~\onlinecite{Lee02a}. The populations and broadening parameters,
as well as the energy differences $\Delta E_{ji}$ and dipole
matrix elements $d_{ij} = ez_{ij}$ relevant for the following discussion
are given in Table~\ref{tab.par}. In addition to the shift due to the
mean-field potential, the energy differences given in Table~\ref{tab.par} 
include also the renormalization due to electron-phonon
and interface roughness scattering.\cite{Lee02a} We neglect impurity scattering
since the doping density per period, $n_e = 2.8 \times 10^{10}$ cm$^{-2}$, 
for this structure is low.

\begin{figure}
\includegraphics[height=5.5cm,keepaspectratio]{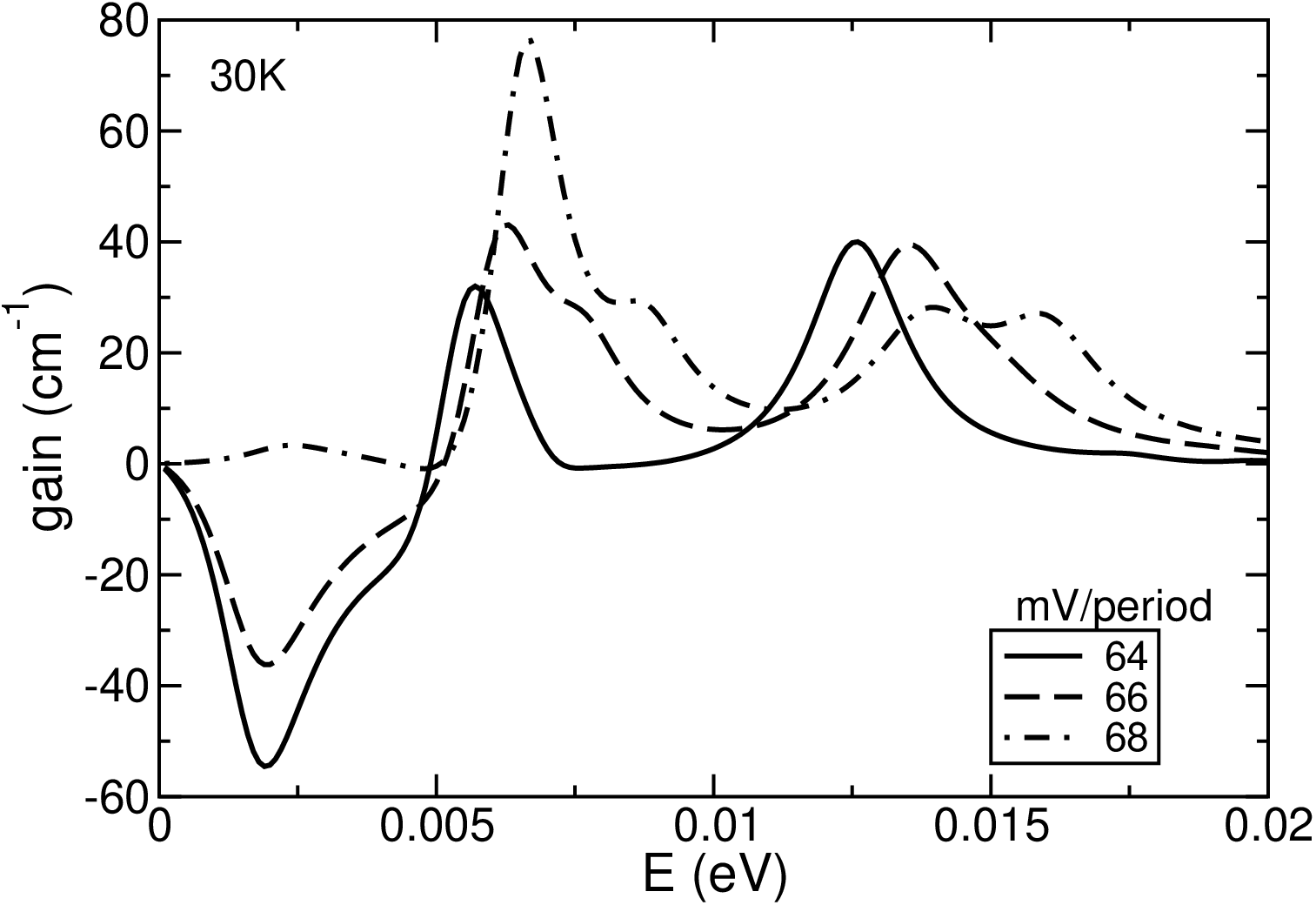}
\caption{Calculated gain spectra at 64, 66, and 68 mV/period.}
\label{fig.spec}
\end{figure}

Figure~\ref{fig.spec} shows the calculated
gain spectra at 30 K for three applied voltages, 64, 66, and 68 mV/period. 
At each of the two lower voltages, two strong gain features are seen:
one lying between 10 and 15 meV, the other between 5 and 8 meV.
There is a blue shift of the gain features as the
bias increases in agreement with the experimental data.\cite{Wil03}
At 68 mV/period, the two gain features merge giving a single broad
gain feature stretching from $\simeq$ 5 to 17 meV.
To see the origin of these features, we look at the contributions
to the spectra from transitions between each pair of levels.

\begin{figure}
\includegraphics[height=10cm,keepaspectratio]{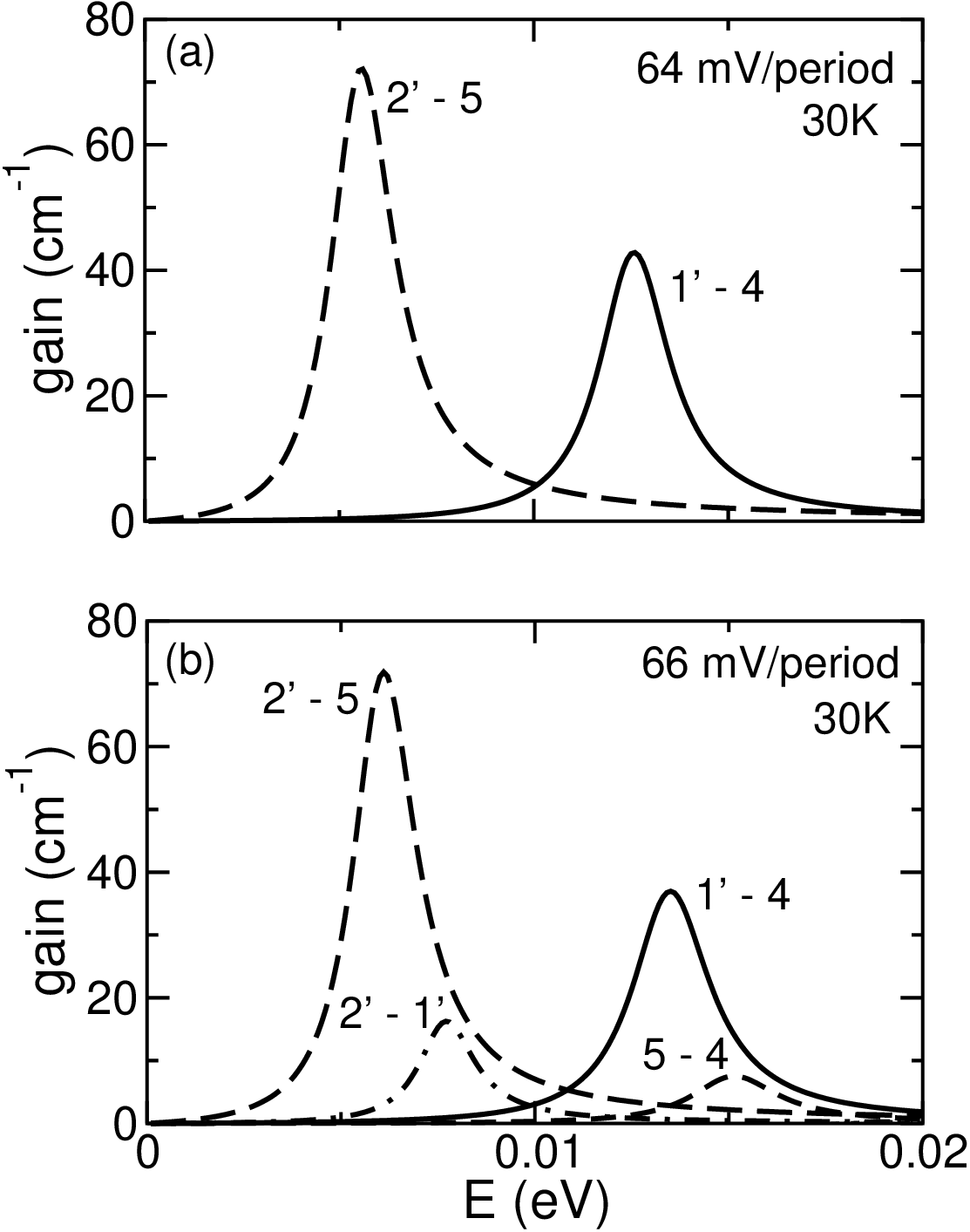}
\caption{Main contributions to  gain spectra. (a) 64 and (b)
66 mV/period.}
\label{fig.spectrans}
\end{figure}

Figure~\ref{fig.spectrans}(a) shows the main transitions contributing
to the gain for the applied voltage 64 mV/period. 
Contrary to the assignment given in Ref.~\onlinecite{Wil03}, we find that 
the strong gain feature at around 12.6 meV is due to the 
transition between the lower collector level (1') and the
lower laser level (4), and not between the upper and lower laser levels.
In fact, at this bias (see table \ref{tab.par}), 
the population in the lower laser level (4) is still
slightly larger than the population in the upper laser level (5),
giving rise to absorption rather than gain between these two levels.
The dipole matrix element between levels 1' and 4 ($|z_{1'4}| = 3.2$ nm) 
is of the same order, if slightly smaller than
that between levels 4 and 5 ($|z_{54}| = 5.14$ nm). 
As importantly, however, there is
a large  population in the lower collector level 1' (in fact, the
most highly occupied level) and this gives rise to a substantial
population inversion between levels 1' and 4.

The strong gain feature at around 5.5 meV arises from the
transition between the upper collector level (2') and the {\em upper}
laser level (5). This transition is favored by the large
dipole matrix element ($|z_{2'5}| = 6$ nm), larger than that between 
levels 4 and 5. The large population in 2' again gives rise to a 
significant population inversion.

Figure~\ref{fig.spectrans}(b) shows the gain contributions at 66 mV/period.
As in the previous case, the main contributions to the gain
arise again from the transitions between levels 1' and 4 (13.6 meV),
and 2' and 5 (6 meV). In addition, there are two smaller gain peaks.
One peak at 15.2 meV can be assigned to the transition between levels 
4 and 5, i.e., the lasing transition considered in Ref.~\onlinecite{Wil03}.
The other small peak at 7.8 meV is assigned to the transition between
the collector levels 1' and 2'. At the highest bias (68 mV/period)
in Fig.~\ref{fig.spec}, the contributions from these 
two latter transitions (4 -- 5 and 1' -- 2')
become stronger leading to the double peaked structure seen at
$\simeq 15$ meV, and the shoulder at $\simeq 8$ meV.
The blue shift of the overall gain features with increasing bias
can therefore be attributed
to the appearance of these secondary peaks, as well as to the Stark
shift mentioned in Ref.~\onlinecite{Wil03}.

In addition to the gain features discussed above, the 
calculations also show a strong gain feature at around 2 meV
at the lower voltage of 48 mV/period. This feature disappeared
as the bias was increased. However, its reappearance is seen again
in the low energy region of  the 68 mV/period spectrum in Fig. \ref{fig.spec}. 
At a higher voltage of 70 mV/period this feature (originating
from the 1' -- 5 transition) becomes stronger.

Thus, a theoretical analysis of this structure shows a
complex behavior of the gain spectra, and their dependence on the
applied voltage. Several transitions contribute to the gain
spectra, with the spectral position and strength of each
contribution depending sensitively on the applied bias.
This suggests the possibility of generating gain
in a spectral region stretching from $\simeq 2$ -- 17 meV, with,
in addition, some ability to tune the wavelengths at which the
gain is enhanced by varying the applied bias. 

We discuss next the robustness of our results to changes in the
parameters used in the calculation.
The values we have used for the broadening parameters $\Gamma_i$ tend to
overestimate the intersubband relaxation rate because $\Gamma_i$ also
includes the effect of intrasubband scattering processes.
We have, however, also used
an alternative approach\cite{Wac02b} to calculate the gain spectra,
in which the linear response of the Green's functions and self-energies
to an optical field are evaluated directly. This approach, which
does not use the Lorentzian 
lineshape function, and hence does not require the parameters $\Gamma_i$,
reproduces the same gain features, in particular, the strong
peaks around 6 meV, and between 10 -- 15 meV. We have also repeated
the calculations at 5 K, and including one extra level (i.e, considering
six levels instead of five), and the main results of the above analysis
are reproduced.
The band structure parameters we used to describe the superlattice
potential [conduction band offset (66\%): 
$\Delta E_c = 0.15$ eV; electron effective masses: $m_e^w = 0.067$
in well, $m_e^b = 0.07945$ in barrier] are extracted from 
Ref. \onlinecite{Vur01}. The resulting theoretical current densities 
are approximately double the experimental values reported in 
Ref. \onlinecite{Wil03}. If, however, we use an 80\% conduction band
offset ($\Delta E_c = 0.1825$ eV) as reported in Ref. \onlinecite{Wil03},
the calculated current densities are reduced, coming in close agreement
with the experimental measurements. The main gain features discussed
above are still present but slightly red-shifted, e.g., for 0.064 mV/period,
the 5.5 meV feature in Fig. \ref{fig.spectrans}(a) 
is shifted down to 4 meV ($\simeq$ 1 THz). 
The height of this gain feature increases by $\simeq 50$\%, 
while the height of the higher frequency gain
feature decreases by $\simeq 25$\%.

We have reported here a detailed analysis of the
gain spectra of a THz quantum cascade structure described in a
recent Letter.\cite{Wil03} The gain spectra exhibits a complicated
behavior and dependence on the applied voltage, with several different
transitions contributing to the gain. Surprisingly, besides the
main lasing transition designated in Ref.~\onlinecite{Wil03}, there
are strong contributions to the gain from neighboring transitions,
particularly transitions between the collector and upper or lower
laser levels. These contributions are in fact larger than 
that of the designed lasing transition. The origin of these
additional contributions to the gain is a sufficiently large
overlap of the wavefunctions in the collector levels with those
of the upper or lower laser levels to give dipole matrix elements comparable
in magnitude to that of the lasing transition, as well as the
large accumulation of population in the collector levels giving
rise to a substantial population inversion with respect to the
upper or lower laser levels. A notable finding is the appearance of
a strong gain feature at around 5 meV, suggesting that this
cascade structure design could also form the basis for a laser
operating at around 1.2 THz ($\lambda\simeq 250$ $\mu$m). 
There was no consideration
or mention of this long wavelength spectral region in 
Ref.~\onlinecite{Wil03}, which focused on the lasing emission around 3.4 THz. 
Thus, the results we report here urge more experimental 
investigations on this interesting structure, both for the prospect
of a THz laser at a longer wavelength than achieved to date for
any other quantum cascade laser structure, and as a test and verification
of our theoretical approach and its predictive power.

This work is supported by the Deutsche Forschungsgemeinschaft.

\clearpage
\begin{table}
\caption{Parameters used in gain calculation for 64 and 66 mV/period. 
$n_i/n_e$ is the fractional
population (total population normalized to 1) in level $i$. $\Gamma_i$
is the broadening parameter (in energy units) for level $i$. 
$\Delta E_{ji}$ is the energy difference between levels $j$ and $i$. 
These parameters are extracted from the quantum kinetics theory.
$d_{ij} = e |z_{ij}|$
is the dipole matrix element between wavefunctions of levels $i$ and $j$.}
\begin{center}
\begin{tabular}{l r @{.} l r @{.} l}
\hline\hline
mV/period  (kVcm$^{-1}$)   &  
\multicolumn{2}{c}{ 64 (12.2) }    & 
\multicolumn{2}{c}{ 66 (12.6) } \\
J (kAcm$^{-2}$)           &  1&6               & 1&8  \\
\hline
$n_1/n_e$                 &  0&35              &  0&26      \\
$n_2/n_e$                 &  0&33              &  0&36      \\
$n_3/n_e$                 &  0&16              &  0&16      \\
$n_4/n_e$                 &  0&08              &  0&096     \\
$n_5/n_e$                 &  0&077             &  0&12      \\
$\hbar/\Gamma_1$ (ps)     &  0&94              &  0&71      \\
$\hbar/\Gamma_2$ (ps)     &  0&99              &  0&81      \\
$\hbar/\Gamma_3$ (ps)     &  0&71              &  0&79      \\
$\hbar/\Gamma_4$ (ps)     &  0&44              &  0&44      \\
$\hbar/\Gamma_5$ (ps)     &  0&5               &  0&55      \\
$E_{54}$ (meV)            & 14&2               & 15&1      \\
$E_{1'4}$ (meV)           & 12&5               & 13&4      \\
$E_{2'5}$ (meV)           &  5&5               &  6&0      \\
$E_{32}$ (meV)            & 39&4               & 39&4     \\
$|z_{54}|$ (nm)           &  5&14              &  4&6      \\
$|z_{1'4}|$ (nm)          &  3&2               &  3&8      \\
$|z_{2'5}|$ (nm)          &  6&1               &  5&9      \\
\hline\hline

\end{tabular}
\end{center}
\label{tab.par}
\end{table}

\clearpage


\end{document}